\documentclass[12pt]{article}
\usepackage{jhep-mod-no-toc}
\usepackage{bm}
\usepackage{amssymb,amsmath,amsthm}
\usepackage{mathrsfs}
\usepackage[utf8]{inputenc}
\usepackage[pdftex]{hyperref}
\hypersetup{colorlinks=true} 
\usepackage{xcolor}
\usepackage{appendix}
\usepackage{graphicx}
\usepackage{dcolumn}
\usepackage{bm}
\usepackage{multirow}
\usepackage{float}
\usepackage[normalem]{ulem}
\usepackage{graphicx}% Include figure files
\usepackage[T1]{fontenc}
\begin{document}
%========================================================
%%--------------------------------------------------------------------------------------------------------------------------------------------
%% very standard definitions
%%------------------------------------------------------------------------------------------------------------------------------------------
\def\d{{\mathrm{d}}}
\def\g{{\mathfrak{g}}}
\def\O{{\mathcal{O}}}
\def\I{{\mathcal{I}}}
\def\Im{{\mathrm{Im}}}
\def\Re{{\mathrm{Re}}}
%-------------------------------------------------------------------------------------------------------------------------------------------
%**************************************
%        New Commands               * 
%**************************************
\newcommand{\PR}[1]{\ensuremath{\left[#1\right]}}
\newcommand{\PC}[1]{\ensuremath{\left(#1\right)}}
\newcommand{\chav}[1]{\ensuremath{\left\{#1\right\}}}
\newcommand{\ve}[1]{\ensuremath{\langle #1\rangle}}
%-------------------------------------------------------------------------------------------------------------------------------------------
\def\lint{\hbox{\Large $\displaystyle\int$}}   %needs \usepackage{amssymb} [large integral]
\def\hint{\hbox{\huge $\displaystyle\int$}}   %needs \usepackage{amssymb} [large integral]
\def\tr{{\mathrm{tr}}}
%-----------------------------------------------------------------------------------------------------
\definecolor{purple}{rgb}{1,0,1}
\newcommand{\red}[1]{{\slshape\color{red} #1}}
\newcommand{\blue}[1]{{\slshape\color{blue} #1}}
\newcommand{\purple}[1]{{\slshape\color{purple} #1}}
%-----------------------------------------------------------------------------------------------------
\parindent0pt
\parskip7pt
%-----------------------------------------------------------------------------------------------------
\allowdisplaybreaks
%-----------------------------------------------------------------------------------------------------
%-----------------------------------------------------------------------------------------------------
\title{\Large{Gravity's universality: The physics underlying Tolman temperature gradients}}
\author{\Large Jessica Santiago$^*$ {\sf and} Matt Visser}
\affiliation{School of Mathematics and Statistics,
Victoria University of Wellington; \\
PO Box 600, Wellington 6140, New Zealand.}
\emailAdd{\{jessica.santiago,matt.visser\}@sms.vuw.ac.nz}
%-----------------------------------------------------------------------------------------------------
\abstract{\ \\
We provide a simple and clear verification of the physical need for temperature gradients in equilibrium states when gravitational fields are present. Our argument will be built in a  completely \emph{kinematic} manner, in terms of the \emph{gravitational red-shift/blue-shift} of light, together with a relativistic extension of Maxwell's two column argument. We conclude by showing that it is the \emph{universality} of the gravitational interaction (the uniqueness of free-fall) that ultimately permits Tolman's equilibrium temperature gradients without any violation of the laws of thermodynamics. 

\medskip\noindent
$^*$ Corresponding author.
		
\medskip\noindent
{\sc Date:} 23 March  2018, 15 May 2018; LaTeX-ed \today.

%\medskip\noindent
%{\sc Length:}  1500 words $\approx$ 125 lines $\approx$ 4 pages... 
%
%\medskip\noindent
%{\sc Date:} BEFORE 1$^{st}$ April...
%

%\medskip\noindent
%\emph{Preprinted as:}  arXiv: 1803.nnnnn [gr-qc]
%
%\medskip\noindent
%\emph{Published as:} 

\medskip\noindent
{\sc Pacs:} 04.20.-q;  04.40.-b; 05.20.-y; 05.70.-a

\medskip\noindent
\vspace{0.5cm}

\leftline{\emph{First prize essay written for the Gravity Research Foundation 2018 Essays on Gravitation.}}

}
%-----------------------------------------------------------------------------------------------------
%\pacs{04.20.-q;  04.40.-b; 05.20.-y; 05.70.-a}
%-----------------------------------------------------------------------------------------------------
\notoc
\maketitle
%-----------------------------------------------------------------------------------------------------
\def\eref#1{(\ref{#1})}
%-----------------------------------------------------------------------------------------------------
%Essays should be 1500 words or fewer excluding abstracts and a small number of equations, diagrams, tables and references. 
%The subject matter may or may not be original research. 
%The essay competition is not intended to replace a research journal where the detailed results of original research are submitted. 
%Essays should not give lengthy detailed mathematical calculations nor detailed descriptions of an experimental setup. 
%Essay ideas should be self-contained and understandable - not dependent on reading other documents.
% SUBMIT BEFORE 1st APRIL
%-----------------------------------------------------------------------------------------------------
%\section{Introduction}
%-----------------------------------------------------------------------------------------------------
\clearpage

Gravity is universal. That, for sure, is something that you will come across when learning general relativity. Einstein's \emph{gedankenexperiment}, with the observer inside an elevator in a sufficiently long free-fall, having time to experience her own limited spacetime, just like Alice falling down the rabbit hole and wondering about the universe, is a really inspiring and surprisingly useful way to do physics. We will apply some of that way of thinking in this essay.

The question we will like to ask and answer in here is whether all the consequences of this universal character of gravity have already been fully explored. We believe that they have not. A major goal of this essay will be to simply and clarify the ideas behind already existent and well known phenomena. Maybe to show a new interpretation and way to see the world, more precisely, thermodynamics. We aim to arrive at the end, to the conclusion that general relativity has a lot to add to classical thermodynamics, in the same way that thermodynamics can change our way to see some gravitational effects. Neither black holes nor horizons nor fire-walls will be present. The systems we will be analyzing are simply boxes full of gases in a pre-defined version of thermal equilibrium.

We can start by defining \emph{thermal equilibrium}.  Textbooks on non-relativistic thermodynamics normally present something along these lines: If a system is in thermal equilibrium, its temperature field $T(t,x_i)$ will be a constant, $t$ being the proper time of the observers at rest with respect to the fluid and $x_i$ the spatial components in the chosen coordinate system. On the other hand, relativistic thermodynamics already understands that this definition is somewhat misleading when gravity is present. 

Relativistic thermal equilibrium in a gravitational field has the interesting feature that the locally measured temperature $T(x_i)$, that is, what a physical thermometer would measure, has a small non-zero spatial gradient~\cite{tolman:1930,ehrenfest:1930} \footnote{Somewhat confusingly, in Tolman's original articles, he uses $T_0$ to denote the locally measured position-dependent temperature --- which we denote $T(x_i)$. In contrast, we reserve $T_0$ for the redshifted temperature, which is spatially constant in thermal equilibrium.}. Indeed
\begin{equation}
{\nabla T(z)\over T(z)} = - {g\over c^2}
\label{E:grad}
\end{equation}
in the flat-earth approximation, with $g$ being the local gravity acceleration and $c$ the speed of light. Near the surface of the Earth this gradient is approximately $10^{-16} \mathrm{m}^{-1}$; certainly negligible in most experimental settings.  
However, the fact that this is non-zero implies that there must be subtle revisions in the non-relativistic versions of the zeroth law, as well as the Clausius version of the second law, and the non-relativistic Fick's law \cite{Santiago:2018}. 

The Tolman result was first established in general relativity~\cite{tolman:1930,ehrenfest:1930}, where in any \emph{static} gravitational field, that is, a spacetime metric which can be put in the block-diagonal form\footnote{In terms of the gravitational potential $N(x)\approx 1+\Phi(x) \approx 1+gz/c^2$.}: 
\begin{equation}
\label{E: metric}
ds^2 =  - N(x_i)^2 \;dt^2 + g_{ab} \;dx^a dx^b,
\end{equation}
one can show
\begin{equation}
\label{E:tolman temp}
T(x_i) = {T_0 \over N(x_i)}; \qquad  T_0 = T(x_i) \, N(x_i).
\end{equation}

In this way, we see that we have two distinct temperature functions. One is the locally measured, position-dependent $T(x_i)$ Tolman temperature. The second, $T_0$, represents the red-shifted temperature.\footnote{\;$T_0$ is the physical temperature at the place where the gravitational potential $\Phi$ is chosen to be zero.
For instance, if we normalize so that $\Phi=0$ at $z=0$, then $T_0$ is just the physical temperature at $z=0$. 
On the other hand, if we normalize so that $N\to1$ at spatial infinity, then $T_0$ is the physical temperature at spatial infinity.} It is this $T_0$ temperature, which is spatially constant in thermal equilibrium, that drives the direction of heat flow in the Clausius version of the second law, and it is gradients in this temperature that drive the modified version of Fick's law:
\begin{equation}
\hbox{(heat flux)} \propto \nabla T_0 = \nabla \left( T(x_i)  N(x_i)  \right). 
\end{equation}

What we want to argue now is that, although extremely well written and didactic, Tolman's work \cite{tolman:1930} creates the impression that the necessary inputs to obtain the temperature gradient are considerably more complex and demanding than they actually are. 
Stripped to its essence, Tolman's argument can be rephrased in a completely \emph{kinematic} manner in terms of the \emph{gravitational red-shift/blue-shift} of light. 

%-----------------------------------------------------------------------------------------------------
\begin{figure}[!h]
	\begin{center}
		\includegraphics[scale=0.44]{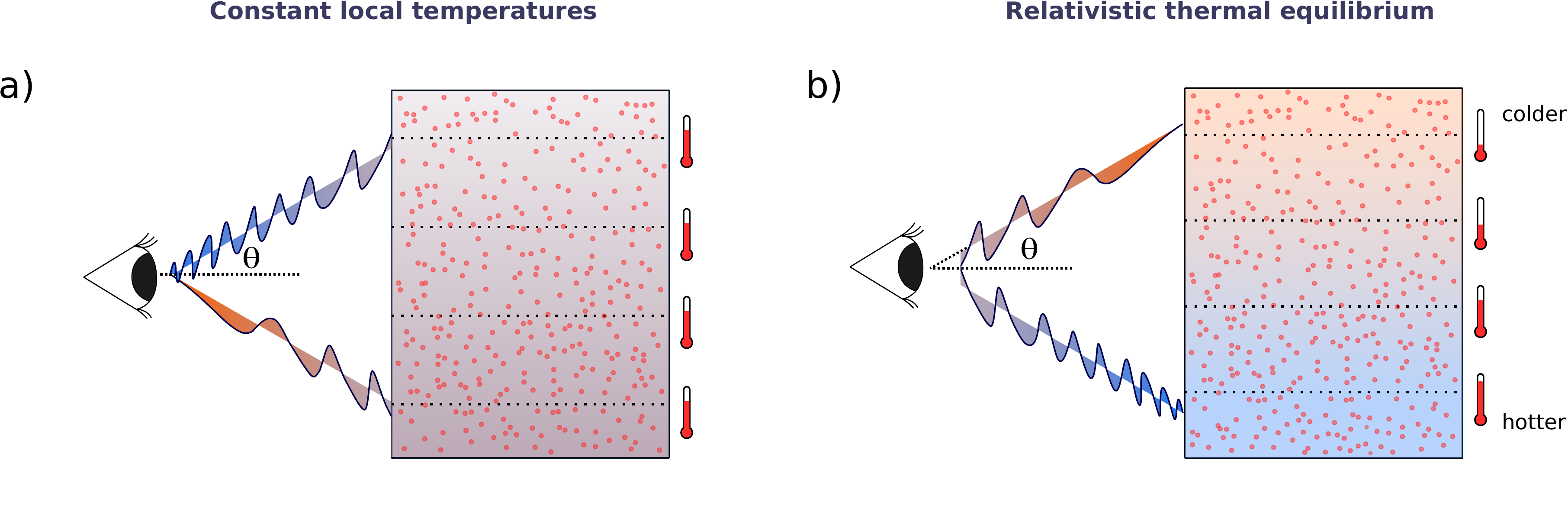}
		\vspace{-0.5cm}
		\caption{\label{F:thermometers2}External observer looking at photons leaking from: a) the box containing the photon gas with constant temperature; and b) the box with the photon gas in relativistic thermal equilibrium.}
	\end{center}	
\end{figure}
%-----------------------------------------------------------------------------------------------------

Let us further explore the meaning of these words. Consider a static observer looking at an angle $\theta$ (with the horizontal) to a photon gas in which all local thermometers measure the same local temperature $T(x_i)$  (see figure~\ref{F:thermometers2}-a). As it is well known, if a photon is travelling along the $\theta$ direction from some distance $r$, it would be coming from a height $z=r \sin\theta$ and would suffer a red-shift/blue-shift of a factor $(1+gz/c^2)$ by the time it finally arrives at the observer. 

On the other hand, photons coming from different gravitational equipotential slices suffer distinct gravitational blue-shifts/red-shifts. 
In this way, if the locally measured temperatures $T(x_i)$ are spatially constant, 
the observer will see not a simple Planck spectrum, but rather a \emph{superposition} of Planck spectra of different temperatures. 

\bigskip
But in this case, the system is not at thermal equilibrium from the point of view of the observer; we have a \emph{reductio ad absurdum}.  The only way to avoid inconsistency is taking into account Tolman's temperature gradients, i.e. to have a position-dependent temperature $T(x_i)=T_0/(1+gz/c^2)\approx T_0 \;(1-gz/c^2)$, since then the gravitational red-shift/blue-shift will \emph{exactly} cancel the gradient factor from the temperature, guaranteeing that all the Planck spectra photons, when seen by the static observer, will be perceived to have the \emph{same} temperature $T_0$ (see figure~\ref{F:thermometers2}-b).

A crucial point in here is that, although it is important that an external observer will perceive a constant temperature when looking at the box, the fact the photons \emph{inside} the box will inevitably suffer blue-shifts/red-shifts when moving around is what actually proves the necessity for temperature gradients in equilibrium states in the presence of gravitational fields.

Now, coming back to gravity's universality, how are these ideas related? The fact is that, without gravity's universal character, Tolman's temperature gradient would indeed violate the second law of thermodynamics. The first to notice this was James Clerk Maxwell \cite{maxwell:1868, maxwell:1902}, as can be seeing in the following argument regarding the equilibrium temperature of a vertical column of gas: 

\begin{quote}
	\textit{``\emph{[...]}	
		if two vertical columns of different substances stand on the same 
		perfectly conducting horizontal
		plate, the temperature of the bottom of each column will be
		the same ; and if each column is in thermal equilibrium of
		itself, the temperatures at all equal heights must be the same. 
		In fact, if the temperatures of the tops of the two columns
		were different, we might drive an engine with this difference of temperature, 
		and the refuse heat would pass down the colder column, 
		through the conducting plate, and up the warmer
		column; and this would go on till all the heat was converted
		into work, contrary to the second law of thermodynamics.''} 
\end{quote}

Its clear from this argument that horizontal (equilibrium) temperature differences cannot exist at equal heights. They will necessarily drive heat fluxes and would allow the construction of a \emph{perpetuum mobile}. In that manner, the vertical temperature gradients, if present at all, must be the same in both columns, regardless the material they are made of. In other words, they must be \emph{universal}. This is a perfectly valid generally correct statement. However, Maxwell's original argument actually has a continuation:

\begin{quote}
	\textit{
	``But we know that if one of the columns is gaseous, its temperature is uniform \emph{[from the non-relativistic kinetic theory of gases]}. Hence that of the other must be	uniform, whatever its material.''}	
\end{quote}

This final conclusion, as we now know, ends up not being valid relativistically\footnote{It is important to point out that, in the non-relativistic limit, Maxwell is actually correct, given that $\nabla T(z) \to 0$ for $c\to\infty$, as can be seen from equation \eqref{E:grad}.}. Given the radiation gas discussion above, we can even rewrite a relativistic version of this final conclusion as:

\begin{quote}
	\textit{But we know that if one of the columns is a photon gas, its temperature must be position dependent, as given by Tolman's relation. Hence that of the other must be position dependent as well, whatever its material.}
\end{quote}

In this way, Tolman's temperature gradient can be proved with simple arguments for any kind of material or physical state of matter.

But, what about the other forces, \emph{e.g.} electromagnetism? Are equilibrium temperature gradients in an electron gas, for example, completely ruled out? Well, yes~\cite{Santiago:2018}. To prove this
simply consider an electron gas, inside a box, submitted to an external electric field. Admit thermal equilibrium. Assume also a very low gas density, so that the force exerted by the external field is much stronger than the interactions between electrons. If a temperature gradient occurs, it will be aligned with the electric field direction. No gravitational field is present.

In case a temperature gradient exists after equilibrium is reached, we can, for example, place next to the just mentioned electron gas a box filled with electrically neutral particles, \emph{i.e.} photons, neutrons, \emph{etc.} Due to its neutrality, it will not interact with the electric field, thus having no reason at all to present a temperature gradient. Continuing the argument on the same lines as Maxwell did, we can see that this situation would indeed generate heat flows, enabling a perpetual motion machine of the second kind. 

The validity of this argument comes from the non-universal character of electric forces. All we needed being two different compositions for the columns, one which reacts to the electric field (charged particles) and one that doesn't. But that machinery can easily be extended to any force which is not universal, allowing us to say:

\emph{Given that temperature gradients created by any force that is not universal (e.g. dependent on charge, mass, spin,...) allow the creation of heat machines that violate the second law, these temperature gradients must not exist. }

Going even further, up to date no force other than gravity seems to act on all sources of mass or energy in the same way regardless, so we can even state this as:

\emph{Gravity is the only force capable of creating temperature gradients in equilibrium states without violating any law of thermodynamics.}

%-----------------------------------------------------------------------------------------------------
\section*{Acknowledgements}
%-----------------------------------------------------------------------------------------------------
JS was supported by a Victoria University of Wellington PhD Scholarship.\\
JS wishes to thank Cesar Uliana-Lima and Uli Zuelicke for helpful discussions.\\
MV was supported by the Marsden Fund, administered by the Royal Society\\
 of New Zealand. 
%-----------------------------------------------------------------------------------------------------
%-----------------------------------------------------------------------------------------------------
%-----------------------------------------------------------------------------------------------------

%-----------------------------------------------------------------------------------------------------
%
%\bigskip
%\centerline{---\ \#\ \#\ \#\ ---}
%
%%-----------------------------------------------------------------------------------------------------
%\begin{thebibliography}{69}
%%-----------------------------------------------------------------------------------------------------
%%-----------------------------------------------------------------------------------------------------
%\end{thebibliography}
%%-----------------------------------------------------------------------------------------------------


\begin{thebibliography}{69}
%-----------------------------------------------------------------------------------------------------

\bibitem{tolman:1930}
R.~C.~Tolman,
  ``On the weight of heat and thermal equilibrium in general relativity,''\\
  Phys.\ Rev.\  {\bf 35} (1930) 904.
  https://doi.org/10.1103/PhysRev.35.904
  %doi:10.1103/PhysRev.35.904
  %%CITATION = doi:10.1103/PhysRev.35.904;%%
  %127 citations counted in INSPIRE as of 06 Mar 2018
 
\bibitem{ehrenfest:1930}
R.~Tolman and P.~Ehrenfest,
  ``Temperature equilibrium in a static gravitational field,''\\
  Phys.\ Rev.\  {\bf 36} (1930) no.12,  1791.
  https://doi.org/10.1103/PhysRev.36.1791
  %%CITATION = doi:10.1103/PhysRev.36.1791;%%
  %125 citations counted in INSPIRE as of 06 Mar 2018
  
 \bibitem{Santiago:2018}
 J.~Santiago and M.~Visser,
  ``Tolman temperature gradients in a gravitational field'',
  arXiv:1803.04106 [gr-qc].
  %%CITATION = ARXIV:1803.04106;%%

\enlargethispage{30pt}
  
\bibitem{maxwell:1868} 
J.~Clerk~Maxwell, ``On the dynamical theory of gases'', \\
The London, Edinburgh, and Dublin Philosophical 
Magazine %and Journal of Science 
{\bf35} (1868) 215--216.



\bibitem{maxwell:1902} 
J.~Clerk~Maxwell, \emph{Theory of heat},
 (Longmans, Green and Company, London, 1902)
 


%\bibitem{frolov} 
%V.~P.~Frolov and A.~Zelnikov, 
%\emph{Introduction to black hole physics}, \\
%(Oxford University Press, Oxford, 2011).
%
%  
%  \bibitem{Israel:1976}
%  W.~Israel,
%  ``Thermo field dynamics of black holes,''\\
%  Phys.\ Lett.\ A {\bf 57} (1976) 107.
%  doi:10.1016/0375-9601(76)90178-X
%  %%CITATION = doi:10.1016/0375-9601(76)90178-X;%%
%  %378 citations counted in INSPIRE as of 07 Mar 2018
%  
%  \bibitem{Abreu:2010a}
%  G.~Abreu and M.~Visser,\\
%  ``Tolman mass, generalized 
%  surface gravity, and entropy bounds,''\\
%  Phys.\ Rev.\ Lett.\  {\bf 105} (2010) 041302
%  doi:10.1103/PhysRevLett.105.041302\\{}
%  [arXiv:1005.1132 [gr-qc]].
%  %%CITATION = doi:10.1103/PhysRevLett.105.041302;%%
%  %27 citations counted in INSPIRE as of 07 Mar 2018
%  
%  
%  \bibitem{Abreu:2010b}
%  G.~Abreu and M.~Visser,
%  ``Entropy bounds for uncollapsed rotating bodies,''\\
%  JHEP {\bf 1103} (2011) 056
%  doi:10.1007/JHEP03(2011)056
%  [arXiv:1012.2867 [gr-qc]].
%  %%CITATION = doi:10.1007/JHEP03(2011)056;%%
%  %7 citations counted in INSPIRE as of 07 Mar 2018
%  
%  \bibitem{Abreu:2010c}
%  G.~Abreu and M.~Visser,
%  ``Entropy bounds for uncollapsed matter,''\\
%  J.\ Phys.\ Conf.\ Ser.\  {\bf 314} (2011) 012035
%  doi:10.1088/1742-6596/314/1/012035
%  [arXiv:1011.4538 [gr-qc]].
%  %%CITATION = doi:10.1088/1742-6596/314/1/012035;%%
%  %7 citations counted in INSPIRE as of 07 Mar 2018
%  
%    \bibitem{Abreu:2011}
%  G.~Abreu, C.~Barcelo and M.~Visser,
%  ``Entropy bounds in terms of the $w$ parameter,''\\
%  JHEP {\bf 1112} (2011) 092
%  doi:10.1007/JHEP12(2011)092\
%  [arXiv:1109.2710 [gr-qc]].
%  %%CITATION = doi:10.1007/JHEP12(2011)092;%%
%  %14 citations counted in INSPIRE as of 07 Mar 2018
%  
%
%  
%  \bibitem{Padmanabhan:2003}
%  T.~Padmanabhan,
%  ``Gravity and the thermodynamics of horizons,''\\
%  Phys.\ Rept.\  {\bf 406} (2005) 49
%  doi:10.1016/j.physrep.2004.10.003
%  [gr-qc/0311036].
%  %%CITATION = doi:10.1016/j.physrep.2004.10.003;%%
%  %395 citations counted in INSPIRE as of 07 Mar 2018
%  
%  \bibitem{Padmanabhan:2010}
%  S.~Kolekar and T.~Padmanabhan,\\
%  ``Ideal gas in a strong gravitational field: 
%  Area dependence of entropy,''\\
%  Phys.\ Rev. {\bf D83} (2011) 064034
%  doi:10.1103/PhysRevD.83.064034\\{}
%  [arXiv:1012.5421 [gr-qc]].
%  %%CITATION = doi:10.1103/PhysRevD.83.064034;%%
%  %19 citations counted in INSPIRE as of 08 Mar 2018
%  
%%  \bibitem{Padmanabhan:2017}
%%  S.~Bhattacharya, S.~Chakraborty and T.~Padmanabhan,\\
%%  ``Entropy of a box of gas in an external gravitational field $-$ revisited,''\\
%%  Phys.\ Rev.\ D {\bf 96} (2017) no.8,  084030
%%  doi:10.1103/PhysRevD.96.084030
%%  [arXiv:1702.08723 [gr-qc]].
%  %%CITATION = doi:10.1103/PhysRevD.96.084030;%%
%  %1 citations counted in INSPIRE as of 08 Mar 2018
%  
%%  \bibitem{Haggard:2013}
%%  H.~M.~Haggard and C.~Rovelli,\\
%%  ``Death and resurrection of the zeroth principle of thermodynamics,''\\
%%  Phys.\ Rev.\ D {\bf 87} (2013) no.8,  084001
%%  doi:10.1103/PhysRevD.87.084001
%%  [arXiv:1302.0724 [gr-qc]].
%%  %%CITATION = doi:10.1103/PhysRevD.87.084001;%%
%%  %17 citations counted in INSPIRE as of 07 Mar 2018
%%  
%%  \bibitem{umov}
%%Nikolay Alekseevich Umov, Selected Works, 1950,
%% (in Russian)
%
%%\bibitem{xxx}
%%\blue{Definitely more references to be added...}
%%
%
%
%  
%%  \bibitem{Padmanabhan:1989}
%%  T.~Padmanabhan,
%%  ``Phase volume occupied by a test particle around an incipient black hole,''
%%  Phys.\ Lett.\ A {\bf 136} (1989) 203.
%%  doi:10.1016/0375-9601(89)90562-8
%%  %%CITATION = doi:10.1016/0375-9601(89)90562-8;%%
%%  %29 citations counted in INSPIRE as of 08 Mar 2018
%
%-----------------------------------------------------------------------------------------------------
\end{thebibliography}
\end{document}